\documentclass[ %
aps, %
pra, %
twocolumn, %
amsmath, %
showpacs, %
a4paper] %
{revtex4-1}

\usepackage[utf8]{inputenc}
\usepackage[T1]{fontenc}
\usepackage{times}
\usepackage[american]{babel}

\usepackage{amsfonts,amsmath,amssymb}
\usepackage{bm}
\usepackage{braket}

\usepackage{graphicx}

\usepackage{hyperref}

\usepackage{color}
\usepackage[export]{adjustbox}

\usepackage[caption=false,labelformat=simple]{subfig}

\newcommand{\E}{\hat{\mathcal E}}
\renewcommand{\P}{\hat{\mathcal P}}
\renewcommand{\S}{\hat{\mathcal S}}

\newcommand{\EE}{\mathcal{EE}}
\newcommand{\Hcal}{\hat{\mathcal{H}}}

\newcommand{\kp}{k_\mathrm{p}}
\newcommand{\omegap}{\omega_\mathrm{p}}
\newcommand{\OmegaEff}{\Omega_\mathrm{e}}

\newcommand{\RB}{R_\mathrm{B}}
\newcommand{\Labs}{L_\text{abs}}
\newcommand{\vgroup}{v_\mathrm{g}}

\newcommand{\abs}[1]{\left\vert #1\right\vert}
\newcommand{\drm}{\mathrm d}
\newcommand{\krm}[1]{\ket{\mathrm{#1}}}

\newcommand{\erf}{\operatorname{erf}}

\newcommand{\sign}{\operatorname{sign}}
\newcommand{\I}{\mathrm i}

\renewcommand{\vec}{\mathbf}
\begin{document}
	\title{Creation and detection of photonic molecules in Rydberg gases}
	\author{M. Moos}
	\email{mmoos@physik.uni-kl.de}
	\affiliation{Fachbereich Physik and Research Center OPTIMAS, Technische Universität Kaiserslautern, 67663 Kaiserslautern, Germany}
	\author{ R. G. Unanyan}
	\affiliation{Fachbereich Physik and Research Center OPTIMAS, Technische Universität Kaiserslautern, 67663 Kaiserslautern, Germany}
	\author{M. Fleischhauer}
	\affiliation{Fachbereich Physik and Research Center OPTIMAS, Technische Universität Kaiserslautern, 67663 Kaiserslautern, Germany}
	\date{\today}
%
% =============== 
\begin{abstract}
We consider the propagation of photons in a gas of Rydberg atoms under conditions of electromagnetically induced transparency, where they form strongly interacting massive particles, termed Rydberg polaritons. Depending on the strength of the van der Waals-type interactions of the atoms either bunching or anti-bunching of photons can be observed when driving the atoms off-resonantly. The bunching is associated with the formation of bound states. We employ a Green's function approach and numerical wave-function simulations to analyze the conditions for the creation and the dynamics of these \emph{photonic molecules} and their interplay with the scattering continuum which can also show photon bunching. Analytic solutions of the pair-propagation problem obtained from a pseudopotential approximation and verified numerically provide a detailed understanding of bound and scattering states. We find that the scattering contributions acquire asymptotically a robust relative phase which can be employed to separate bound-state and scattering contributions by a homodyne detection scheme.
\end{abstract}
% =============== 
%
\maketitle

% ================================
\section{Introduction}
% ================================
%
Rydberg gases are of great interest in quantum optics as they enable to mediate strong and long-range nonlinearities between photons. The van der Waals-type interactions between Rydberg states \cite{Saffman2010} in a gas of three-level atoms can be used to create strong photon-photon interactions for light fields coupling to the atomic medium in an scheme of electromagnetically induced transparency (EIT) \cite{Fleischhauer2005,Pritchard2010}. Moreover, due to the long-range nature of the Rydberg-Rydberg interactions, also the photon-photon interactions are long-range which makes this setup a promising candidate for creating and analyzing interesting many-body states with applications ranging from quantum computation \cite{Tiarks2014,Gorniaczyk2014,Maxwell2013,Distante2016,Hofmann2013} to quantum simulations \cite{Hafezi2013,Ningyuan2016,Zhang2013,Jachymski2016,Maghrebi2015}.

\begin{figure}
	\centering
	\quad
% ===========
	\subfloat[Level scheme]{
		\includegraphics[width=.25\columnwidth,valign=c]{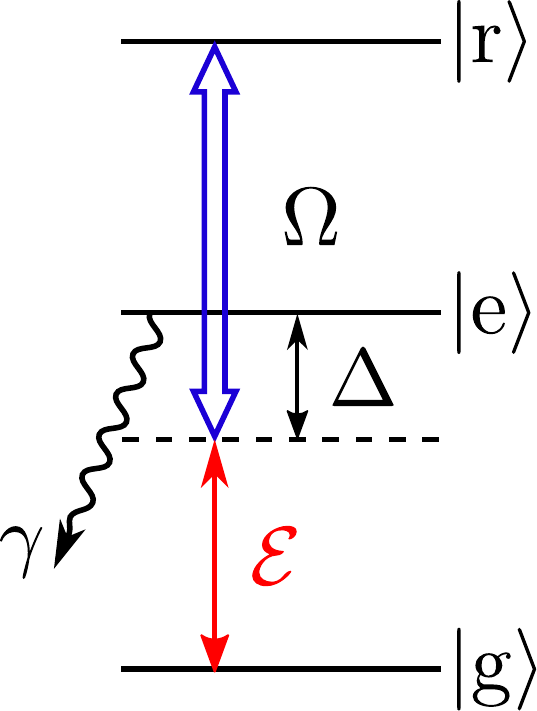}	
		\vphantom{\includegraphics[height = .465\columnwidth,valign=c]{setupHomodyneDetection.pdf}}
		\label{fig:Figure1:a}
	}
% =========== 
	\quad
% ===========
	\subfloat[Detection setup]{
		\quad
		\def\svgwidth{0.5\columnwidth} 				
		\adjustbox{valign=c}{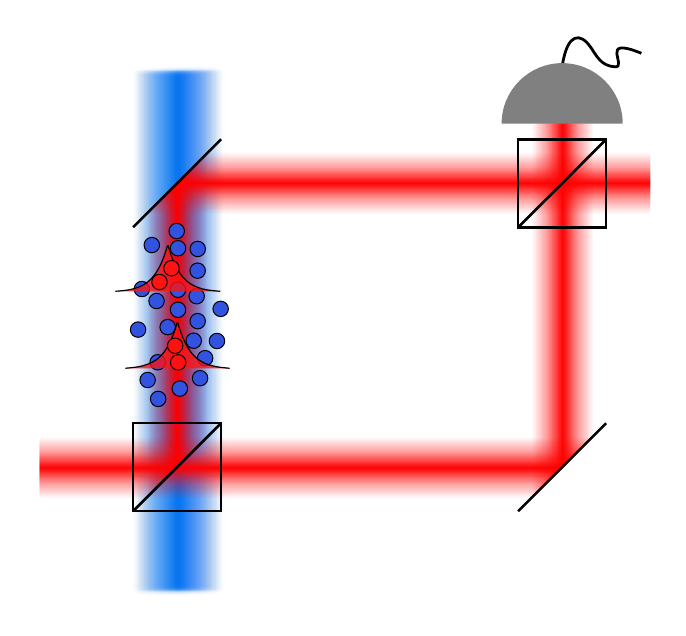}
		\label{fig:Figure1:b}	
	}
% ===========
	\\
% ===========
	\subfloat[$\abs{\mathcal{EE}}^2$ in a.u., $ \xi=0.2 $]{
		\includegraphics[width = 0.47\columnwidth]{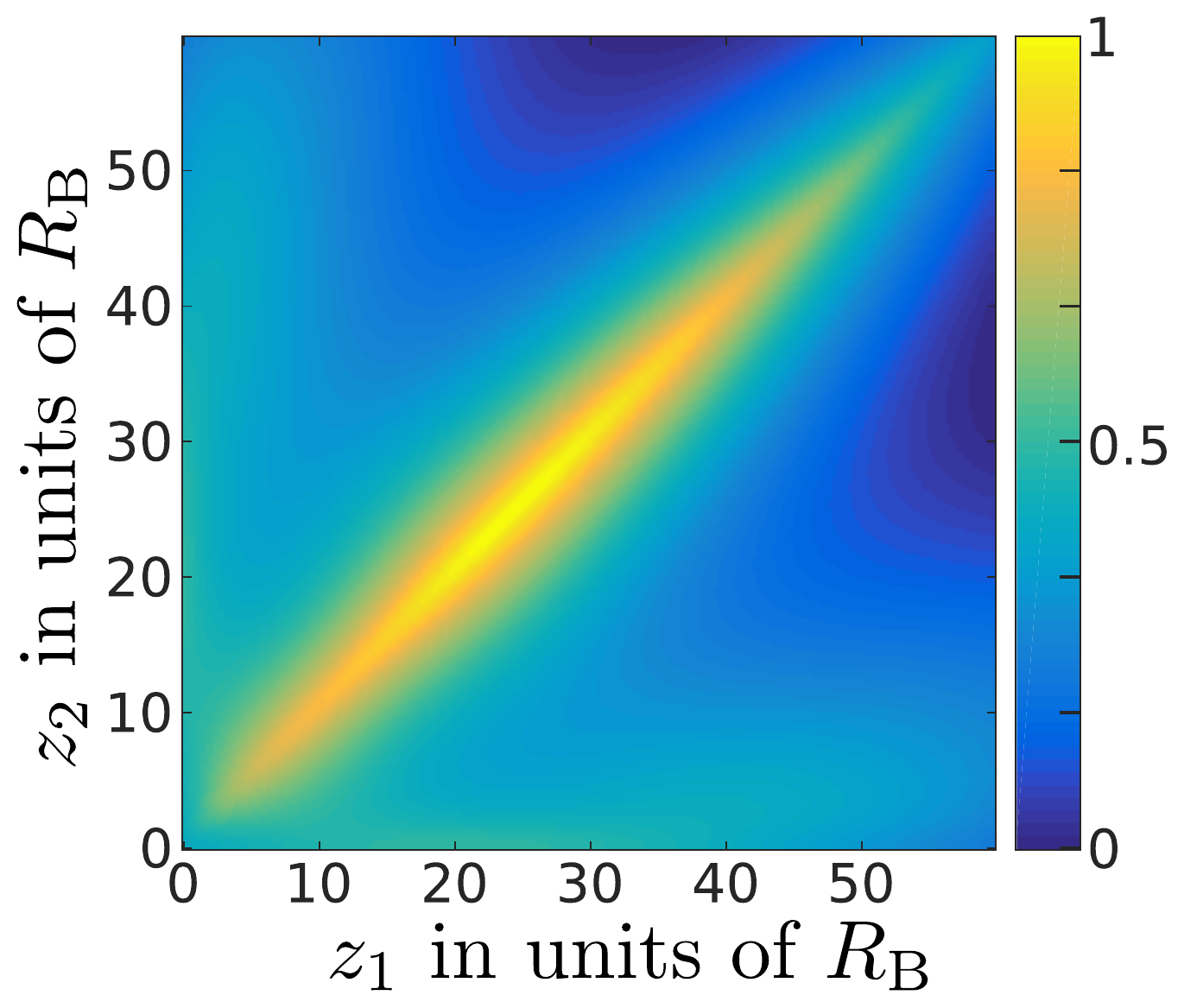}
		\label{fig:Figure1:c}
	}
% ===========
	\hfill
% ===========
	\subfloat[$\abs{\mathcal{EE}}^2$ in a.u., $ \xi=2 $]{
		\includegraphics[width = 0.47\columnwidth]{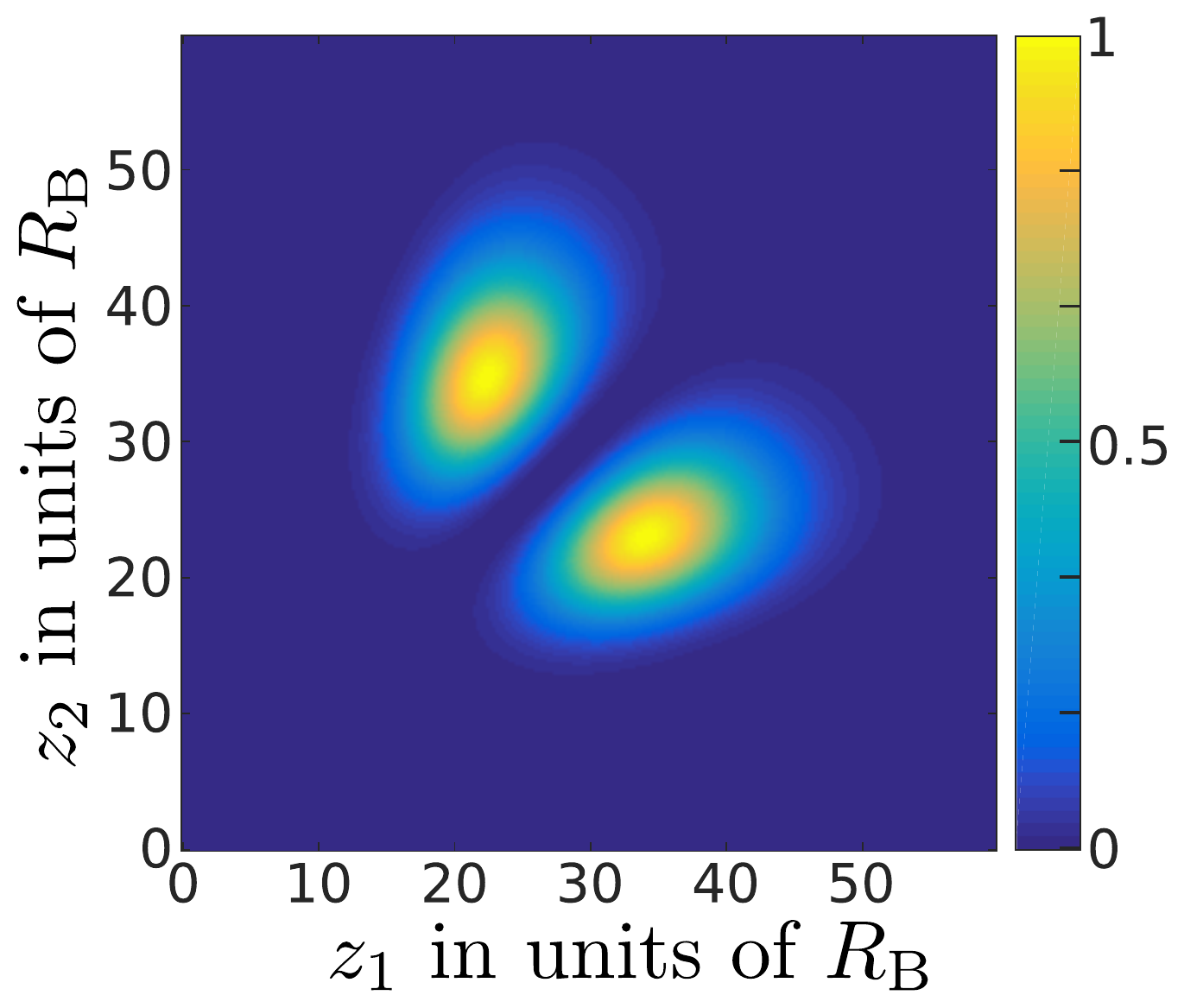}
		\label{fig:Figure1:d}
	}
% ===========
%
	\caption{
	(a) Sketch of atomic coupling scheme in ladder-type EIT setup, consisting of states $\krm{g}, \krm{r}$ and the intermediate state $\krm{r}$ that is subject to spontaneous decay. The probe field $\E$ and the control field $ \Omega $ drive the transitions $ \krm{g}\leftrightarrow\krm{e} $ and $ \krm{e}\leftrightarrow\krm{r} $, respectively.
	(b) Experimental setup for homodyne detection to filter bound state and scattering state components.
	(c) and (d) Numerical simulation of the two-photon wave-function $\EE(z_{1},z_{2},t)=\bra{0}\E(z_{1})\E(z_{2})\ket{\phi(t)}$ inside a three-level atomic medium after propagating from free space into the medium with boundary at $ z=0 $. Depending on the ratio $ \xi=\RB/\Labs $, the two-photon wavefunction has a very different spatial structure.	Here, $ \RB $ denotes the Rydberg blockade distance and $ \Labs $ the off-resonant absorption length in absence of EIT, as defined in section~\ref{sec:modelAndTwoExcitationEquations}.
	 In (c) we show the result for weak interactions, $ \xi=0.2$, where bunching can be observed. For strong interactions, there is an anti-bunching of photons, as shown in (d), where $ \xi=2 $.
	}
\label{fig:FigureOne}
\end{figure}

In an EIT setup photons travel as massive quasi-particles, so-called dark-state polaritons (DSPs) with group velocities much smaller than the vacuum speed of light \cite{Fleischhauer2005}. If the EIT coupling involves atomic Rydberg states (see Fig.\ref{fig:Figure1:a}), interactions between Rydberg atoms are transferred to polaritons. In particular, it has been proposed and observed experimentally that repulsive Rydberg interactions lead to an avoided volume of photons for small distances, i.e., anti-bunching \cite{Gorshkov2011,Peyronel2012a}. On the other hand also bunching of photons has been observed, when driving the atoms off-resonantly \cite{Firstenberg2013}. This can result from photonic bound states ("photonic molecules"), but also from bunched continuum components (scattering states). The formation of bound states requires an interplay between interactions and dissipation.
Numerical simulations show that bunching can only be observed in the regime of weak to moderate polariton interactions, quantified by a small optical depth per blockade, $\xi \ll 1$, which will be defined later-on, see Fig.~\ref{fig:Figure1:c}--\subref*{fig:Figure1:d}. In the present paper we investigate the properties of photonic bound states, analyze conditions for their formation, and discuss possible ways to distinguish bound- and continuum-state contributions.
%

% ==============================
% ==============================

Specifically, we apply a Green's function approach to model one-dimensional systems of interacting Rydberg polaritons and investigate the creation of states leading to the bunching and the dynamics at large times. For short times, bound- and scattering-state contributions are equally important to explain photon bunching. The bound state contribution decays exponentially due to losses, and thus bunching at large times comes solely from scattering states. 

We find that the scattering states have a phase which depends only on the ratio of the probe field detuning and the decay rate of the excited atomic state. Therefore, this phase is very robust and can be used to separate bunched photons resulting from bound states and scattering states by homodyne detection, see Fig.\ref{fig:Figure1:b}.

% ================================
\section{Model and two-excitation equations}
\label{sec:modelAndTwoExcitationEquations}
% ================================
% ================
\subsection{Model}
% ================
%
We consider a quantized probe field $\E$ propagating under conditions of EIT in a medium consisting of $N$ atoms with three levels driven by two optical fields in a ladder scheme as illustrated in Fig.~\ref{fig:Figure1:a}.

The atomic ground state $\krm{g}$ and the excited state $\krm{e}$ are coupled by the quantized probe field $\E(\vec r ,t)$ with carrier frequency $\omega_{\mathrm{p}}$ and wave vector $\vec{k}_{\mathrm{p}}$. The probe field is detuned from the atomic transition by the single-photon detuning $\Delta=\omega_\mathrm{eg}-\omega_{\mathrm{p}}$.  Furthermore, a classical control field with frequency $\omega_{\mathrm{c}}$ drives the transition $ \krm{g}\leftrightarrow\krm{e} $ with Rabi frequency $\Omega$ and detuning $\Delta_{\mathrm{c}}$, which is chosen such that the resulting
two-photon detuning vanishes, i.e., $\delta=\Delta+\Delta_{\mathrm{c}}=0$. The intermediate state is subject to spontaneous decay with rate $\gamma$. The atoms are described by spin flip operators $\hat{\sigma}_{\mu\nu}^{i}=\ket{\mu}_{ii}\bra{\nu}$ and interact via the van der Waals potential $V(\vec r)=C_{6}/\abs{\vec r}^6$ in the level $\krm{r}$. Assuming a homogeneous distribution of atoms we can
describe them by coarse-grained continuous operators $\hat{\sigma}_{\mu\nu}(\vec{r})$, which for negligible atomic saturation are bosonic fields $\S(\vec r)=\hat{\sigma}_{\mathrm{gr}}(\vec r)$ and $\P(\vec r)=\hat{\sigma}_{\mathrm{ge}}(\vec r)$.

Finally, from the atom-field coupling Hamiltonian in rotating wave approximation and Maxwell's equations we obtain in linear response in $g\E$ the paraxial Maxwell-Bloch equations,
\begin{align}
	\I\tfrac{\partial}{\partial t}\E(\vec{r})  
	&  =-\I c\tfrac{\partial}{\partial z}\E(\vec{r})-\tfrac{c}{\abs{\vec{k}_{\mathrm{p}}}}
	\nabla_{\!\perp}^2\E(\vec{r})-g\P(\vec{r}),
	\nonumber\\
	\I\tfrac{\partial}{\partial t}\P(\vec{r})  
	&  = -\I\Gamma\P(\vec{r})-\Omega\S(\vec{r})-g\E(\vec{r})+\hat{F}_{\mathrm{ge}},\\
	\I\tfrac{\partial}{\partial t}\S(\vec{r}) 
	&  = -\Omega\P(\vec{r})+\int\!\drm\vec{r}'\, V(\vec{r}-\vec{r}' ) \S^{\dagger}(\vec{r}' )\S(\vec{r}' )\S(\vec{r}) ,
	\nonumber
\end{align}
where we defined the complex detuning $\Gamma=\gamma+\I\Delta$ and the coupling strength $g=\wp\sqrt{n\omega_{\mathrm{p}}/2\hbar\epsilon_{0}}$ with $\wp$ being the dipole moment of the $ \ket{\mathrm{g}}\leftrightarrow\ket{\mathrm{e}} $ transition, and $n$ being the atomic number density. $\hat{F}_{\mathrm{ge}}$ is a Langevin noise operator~\cite{Louisell1973}, which we introduced to preserve the commutation relations. Under EIT driving conditions the occupation of the level $ \ket{\mathrm e} $ stays small, and thus the Langevin noise can be neglected.

As shown in \cite{Moos2015}, in experimentally relevant situations the interaction can be described by a one-dimensional model and we can neglect the transverse kinetic energy $\tfrac{c}{k_{\mathrm{p}}}\nabla_{\perp}^{2}\hat{\mathcal{E}}$. Finally, assuming the time evolution being slow on the time scale set by the complex detuning $ \abs{\Gamma} $, we adiabatically eliminate the optical polarization  $\P$, leading to
\begin{equation}
\begin{aligned}
	\I\tfrac{\partial}{\partial t}\E(z)  
	& = -\I c\tfrac{\partial}{\partial z}\E(z) - \I\tfrac{g^{2}}{\Gamma}\E(z)- \I\tfrac{g\Omega}{\Gamma}
	\S(z),\\
	\I\tfrac{\partial}{\partial t}\S(z)  
	&  =-\I\tfrac{\Omega^{2}}{\Gamma}
	\hat{\mathcal{S}}(z)-\I\tfrac{g\Omega}{\Gamma}\hat{\mathcal{E}}(z)\\
	&\quad\qquad +\int\drm z'\, V(z-z')
	\S^{\dagger}(z' )\S(z')\S(z),
\end{aligned}
\label{eq:OperatorEquation}
\end{equation}
which is a set of coupled nonlinear integro-differential equations for the operators $\E$ and $\S$.

% ===================================
\subsection{Dark-state polaritons}
% ===================================
%
Let us first briefly summarize the description of the noninteracting limit, i.e., $ V(z)\equiv0 $, which also applies to the case of a single photon propagating through the Rydberg medium. In this case, Eqs.~\eqref{eq:OperatorEquation} form a set of linear equations that can be expressed as
\begin{equation}
	\label{eq:eomAndHamiltonianNoninteracting}
	\I\frac{\partial}{\partial t}\begin{pmatrix}
		\E\\
		\S
	\end{pmatrix}
	= \hat H_0
	\begin{pmatrix}
		\E\\ \S
	\end{pmatrix},\quad
	 \hat H_0 = -\I
	\begin{pmatrix}
		c\frac{\partial}{\partial z} +\frac{g^2}{\Gamma} & \frac{g\Omega}{\Gamma}\\
		\frac{g\Omega}{\Gamma} & \frac{\Omega^2}{\Gamma}
	\end{pmatrix}.
\end{equation}
%
% MM: Our notation with $ \hat H,\Hcal $ is not that consistent...
The eigenmodes of the Hamiltonian $ \hat H_0 $ in the long-wavelength limit ($ k\approx 0 $) correspond to quasi-particles composed of light and matter excitation, the so-called dark- and bright-state polaritons which can be written as $ \hat{\psi}_\mathrm{d} = -\cos\theta\E+\sin\theta\hat{\mathcal S} $ and $ \hat \psi_\mathrm{b} = \sin\theta\E+\cos\theta\hat{\mathcal S} $, respectively, see, e.g., \cite{Fleischhauer2005}. Here the mixing angle $ \theta $ is defined by $ \tan\theta=g/\Omega $. Treating the momentum $ k $ perturbatively one finds that the dark-state polariton propagates lossless with the group velocity $ \vgroup=c\cos^2\theta $. Furthermore, it forms a quasi-particle with an effective mass $ m $. The mass is approximately $ m\approx (2\vgroup\Labs)^{-1} $ under slow-light conditions and in an off-resonant driving scheme, where $ \Labs=\abs{\Delta}c/g^2$ is the off-resonant optical depth,  In contrast, the bright-state polariton propagates with velocity $ c\sin^2\theta \approx c $ and is subject to losses with the rate $ \gamma\OmegaEff^2/\abs{\Delta}^2 $, where the effective Rabi frequency is defined by $ \OmegaEff^2 = g^2+\Omega^2 $. 

For large separations between excitations the Rydberg-Rydberg interaction can be included as a perturbation, see, e.g., \cite{Otterbach2013}. However, this approach does not capture bound states and is thus not applicable in general. Instead the full scattering problem has to be considered as was done in \cite{Bienias2014}; see also \cite{Gullans2016}.

%
% =========================================
\subsection{Effective model for two excitations}
\label{sec:IIC}
% =========================================
%
%
To analyze the dynamics of interacting excitations we now consider the time evolution of two particles, which can be done by using wave functions $\EE(z_{1},z_{2},t)=\bra{0}\E(z_{1},t)\E(z_{2},t)\ket{\phi}$, and analogously defined components $\mathcal{ES},\mathcal{SE}$ and $\mathcal{SS}$ that can be combined into the four-component vector $\bm{\Psi}_2=(\mathcal{EE}, \mathcal{ES}, \mathcal{SE}, \mathcal{SS})^{T}$. The time evolution of $ \bm{\Psi}_2 $ in real space is governed by the equation 
\begin{equation}
	\label{eq:eom2photonWaveFunction}
	\I\frac{\partial}{\partial t}\bm{\Psi} = 
	\left\{
	\Hcal_0(z_1,z_2)
	+
	V(z_1-z_2)\hat{\mathrm{P}}_{\mathcal{SS}}
	\right\}\bm{\Psi}
\end{equation}
with $\Hcal_0 = \hat H_0(z_1)\otimes\openone_2+\openone_2\otimes \hat H_0(z_2)$. The operator
$ \hat{\mathrm{P}}_{\mathcal{SS}}=\ket{\varphi_4}\bra{\varphi_4} $  denotes the projector onto the $ \mathcal{SS} $-component of the wave function, i.e., two Rydberg excitations, with $ \ket{\varphi_4} =(0,0,0,1)^{T} $.
This equation can be integrated numerically to find the time evolution of a two-photon wave packet. In particular we simulate the time evolution starting in free space and propagating according to equation \eqref{eq:OperatorEquation} through a sharp boundary. We find qualitatively a very different behavior inside the medium depending on the strength of the interaction potential $ V(z_1-z_2) $ as can be seen in Figs.~\ref{fig:Figure1:c} and \ref{fig:Figure1:d}. In the weakly interacting regime, Fig.~\ref{fig:Figure1:c},
we find a bunching of photons, while in the strongly interacting regime, Fig.~\ref{fig:Figure1:d},
 the photons avoid a volume given by $ \abs{z_1-z_2}< \RB $, with $R_\mathrm{B}=\left(\abs{\Delta}C_6/2\Omega^2\right)^{1/6}$ being the off-resonant blockade radius .

To gain analytical insight into these observations, we employ a Green's function approach to solve the time evolution of the two-photon wave-function, Eq.~\eqref{eq:eom2photonWaveFunction}, similar to \cite{Bienias2014}. We transform to center-of-mass- and relative  coordinates of the two excitations, $R=\tfrac{1}{2}(z_{1}+z_{2})$ and $r=z_{1}-z_{2}$, respectively. Subsequently, we perform a Fourier transform with respect to the center of mass $R$ according to $ f(R) = \int\drm K\,  \mathrm{e}^{iKR}{\tilde f}(K) $. Specifically, we consider the initial state
\begin{equation}
	\vec{\Psi}(K,r,0)=f(K,r)\ket{\varphi_1} ,
	\label{eq:initial_state_vector}%
\end{equation}
where $\ket{\varphi_1} =(1,0,0,0)^{T}$, i.e., we assume that only the photonic component is present at the beginning of the evolution. Our calculation can easily be generalized to 
other initial states.
Furthermore, we restrict ourselves to the case of negative single-photon detuning, $ \Delta<0 $. The solution for positive detuning can be derived straightforwardly.

We are interested in the (asymptotic) behavior of the amplitude $ \EE(K,r,t) $  at large times. In this limit, the low-frequency contributions are the dominant ones (see Appendix \ref{sec:appendix_a} for more details) and by simple algebraic calculations one obtains for the two-photon amplitude
\begin{equation}
	\EE(K,r,t) =
	\frac{\cos^{4}\theta}{2\pi \I}\iint \drm\omega\, \drm r'\,
	\mathrm{e}^{-\I\omega t} G(r,r',\omega) f(K,r' ).
	\label{eq:photonicEvolutionAsymptotic}
\end{equation}
Here the Green's function $G(r,r',\omega) $ is the solution of the integral equation
\begin{multline}
	G(r,r',\omega) = G_0(r,r',\omega)\\
	-\sin^{4}\theta\int	\drm r''\, G_0(r,r'',\omega) W(r'') G(r'',r',\omega),
	\label{eq:GreenFunctionIntegralEquationSmallFreq}
\end{multline}
where $ W(r) $ denotes an effective potential which is defined by Eq.~\eqref{eq:effective_potential}.

Under the condition that $ \Im\{\sqrt{2m(\omega-\vgroup K)}\}>0 $, the free Green's function $G_{0}$ in Eq.~\eqref{eq:GreenFunctionIntegralEquationSmallFreq} has the coordinate representation
\begin{equation}
	G_0(r,r',\omega) =
	-\frac{
		\exp\{\I\sqrt{2m(\omega-\vgroup K)}|r-r'|\}
		}{
		2\I\sqrt{2m(\omega-\vgroup K)}
		}.
	\label{eq:freeGreensFunction}
\end{equation}

In the low-energy regime, $|\omega| \ll \Omega^2/\vert \Gamma\vert$, i.e., for frequencies well inside the EIT window the Green's function $G$ describes the evolution of a particle with an effective Hamiltonian
\begin{equation}
	\Hcal_\mathrm{eff} =
	-\frac{1}{2m}\frac{\mathrm{d}^2}{\mathrm{d}r^2}+W(r) \sin^{4}\theta .
	\label{eq:Schroedinger_Hamiltonian}%
\end{equation}
The complex mass is given by
\begin{equation}
	m = \I \frac{g^2}{4c\Gamma\vgroup}= \frac{\sign(\Delta)}{4 \vgroup \Labs} \left(1+i\frac{\gamma}{\Delta}\right)
	\label{eq:reduced_mass}
\end{equation}
and the effective interaction potential reads  
\begin{equation}
W(r)\equiv\frac{V(r)}{1+\alpha V(r)},\qquad \alpha = \frac{i\gamma -\Delta}{2\Omega^2}.
\label{eq:effective_potential}
\end{equation}
In the limit of slow light $ g\gg\Omega $ and large single photon detuning $ |\Delta|\gg\gamma $ 
the effective mass reduces to the simpler expression $ m\approx \sign(\Delta)(4\vgroup\Labs)^{-1} $, which coincides with the results derived in \cite{Zimmer2008}.
Likewise the coefficient $\alpha$ simplifies to $\alpha \approx -\Delta/2\Omega^2$. 

In the following we assume slow-light conditions and set $\sin^2\theta \approx 1$.

% ==============================
\begin{figure}
	\centering
	\subfloat[$ \Delta=8\gamma>0 $]{
%	\includegraphics[width = 0.5\columnwidth]{figs/effectivePotentialPosDelta.pdf}
%		\tikzset{external/force remake}
%		\setlength\fheight{0.3\columnwidth} 
%		\setlength\fwidth{0.4\columnwidth}
%		\input{figs/effectivePotentialPosDelta.tikz}	
		\includegraphics[]{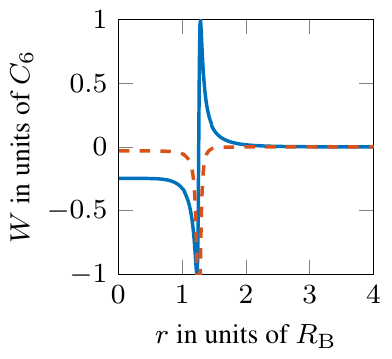}
		\label{fig:effectivePotential:Positive}
	}
	\subfloat[$ \Delta=-8\gamma<0 $]{
%	\includegraphics[width = 0.5\columnwidth]{figs/effectivePotentialNegDelta.pdf}
%		\tikzset{external/force remake}
%		\setlength\fheight{0.3\columnwidth} 
%		\setlength\fwidth{0.4\columnwidth}
%		\input{figs/effectivePotentialNegDelta.tikz}
		\includegraphics[]{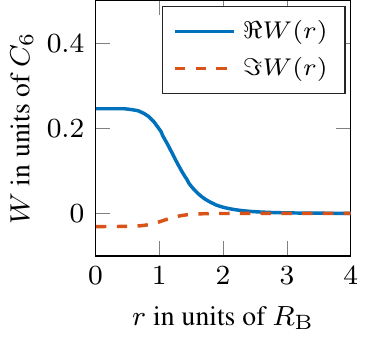}
		\label{fig:effectivePotential:Negative}
	}	
	\caption{Real (solid blue lines) and imaginary part (dashed red lines) of effective potential $ W(r) $, defined in Eq.~\eqref{eq:effective_potential} for (a) positive and (b) negative single-photon detuning $\Delta=\pm8\gamma$ and $ \Omega = \gamma $.}
	\label{fig:effectivePotential}
\end{figure}
% ==============================

In Fig. \ref{fig:effectivePotential} we show the effective potential for positive and negative detuning. For distances larger $ \RB $ the potential decays like the bare van der Waals potential, for small distances the potential becomes flat. We can interpret the effective potential as complex susceptibility of a single photon in the presence of a fixed Rydberg excitation at the origin resulting in a space dependent two-photon detuning \cite{OtterbachThesis,Gorshkov2011}. 
\begin{equation}\label{eq:susceptibiltiy}
	\chi(r) = \chi'+\I\chi'' = -\I\frac{g^2}{\Omega^2} W(r).
\end{equation}

Note that the complex mass \eqref{eq:reduced_mass} always has a positive imaginary part, effectively describing the (small) polariton losses due to spontaneous decay of the intermediate level $ \ket{\mathrm{e}} $, while the sign of its real part can be tuned depending on the sign of  the single-photon detuning $ \Delta $. 
The product of the real part of the effective potential and the effective mass is always negative at small distances, suggesting the existence of bound states, independent of the sign of  $ \Delta $.

%============================================
\section{Weakly bound states - photonic molecules}
% ===========================================
%
For the interacting problem we have to solve Eq.~\eqref{eq:GreenFunctionIntegralEquationSmallFreq} for the Green's function. In the far-detuned limit, when $\left\vert \Delta\right\vert \gg \gamma$, the Green's function $G(r,r',\omega) $ can be written as a sum,
\begin{equation}
	G(r,r',\omega)  
	=
	\sum\limits_{n=1}^{N}
	\frac{\psi_{n}(r)  \psi_{n}^{\ast}(r')}{\omega-E_{n}}
	+
	\int\drm E\,\frac{\psi_{E}( r)  \psi_{E}^{\ast}(r')}{\omega-E},
	\label{Spectral_Representation_Green_Function}
\end{equation}
of bound eigenstates, denoted by $\psi_{n}(r)$, and continuum eigenstates, denoted by $\psi_E(r)$,
The binding energies of the molecular states are in general complex and increase with 
the optical depth per blockade distance, $ \xi\equiv \RB/\Labs $.

A sufficient condition for the existence of bound states $ \psi_n $ in the spectrum of this Hamiltonian is \cite{Simon2005}
\begin{equation}
	\int_{-\infty}^{\infty}\drm r\,m W(r) <0.
	\label{eq:condition_bound_states}
\end{equation}
We note that in our case of negative single-photon detuning, the product of $ m $ and $ W(r) $ is negative, and thus this condition is met.

% =================================================
\subsection{Properties of bound states}
\begin{figure}
	\centering
	\subfloat[Spectrum of effective Hamiltonian]{
%		\tikzset{external/force remake}
%		\setlength\fheight{0.4\columnwidth} 
%		\setlength\fwidth{0.69\columnwidth}
%		\input{figs/effectiveHamiltonianSpectrum.tikz}
		\includegraphics[]{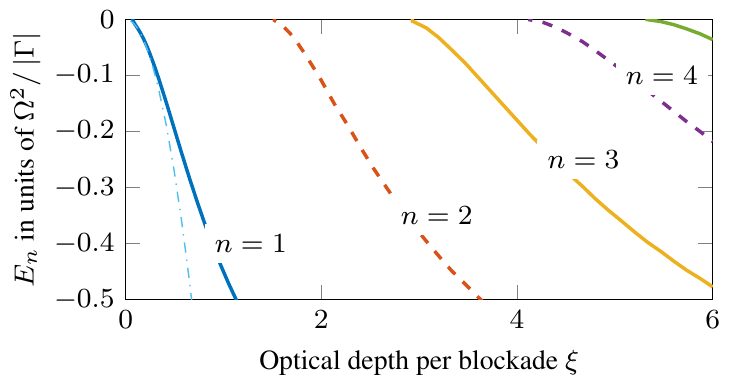}		
		\label{fig:singleBoundStateAndSpectrumHeff:Spectrum}
	}\\
	\subfloat[Bound state and exponential function]{
%		\tikzset{external/force remake}
%		\setlength\fheight{0.4\columnwidth} 
%		\setlength\fwidth{0.7\columnwidth}
%		\input{figs/boundStateAndExponential.tikz}
		\includegraphics[]{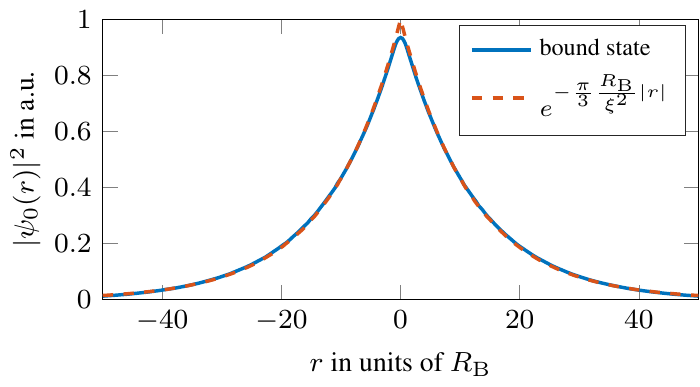}
		\label{fig:singleBoundStateAndSpectrumHeff:BoundState} 
	}
	\caption{ %
	Results of numerical diagonalization of the effective Hamiltonian $ \Hcal_\mathrm{eff} $, Eq.~\eqref{eq:Schroedinger_Hamiltonian} for a system of finite length with periodic boundary conditions.
	(a)	Bound-state energies $ E_n $ of in dependence of interaction strength, respectively optical depth per blockade volume $ \xi $. The (light blue) dashed-dotted line is the approximate solution $ E_0\approx -\tfrac{\pi^2}{9}\xi^2 $, that will be derived in section \ref{sec:boundStatesAndContinuumStates}. We restrict the plot to energies larger $ -\Omega^2/2\abs{\Gamma} $, as the effective Hamiltonian is only valid for small energies.
	(b) First bound state $ \psi_0 $ in comparison to an exponential function in the weakly interacting regime with $ \xi = 0.2 $, where we adjusted the amplitude of the bound state to fit the exponential. 
	}
	\label{fig:singleBoundStateAndSpectrumHeff}
\end{figure}

% =================================================
%
%
The bound eigenstates $ \psi_n(r) $ can be computed by numerical diagonalization of the effective Hamiltonian, Eq.~\eqref{eq:Schroedinger_Hamiltonian}, for discretized spatial coordinates on a finite spatial interval. This allows us to get an approximate spectrum of the bound eigenstates as a function of the optical depth per blockade distance, $ \xi $, which is shown in Fig.~\ref{fig:singleBoundStateAndSpectrumHeff:Spectrum}. For sufficiently small $\xi$ only a single bound state exists and with increasing $ \xi $ the number of bound eigenstates grows, as does their energy.
As the effective Hamiltonian is only applicable in the regime of small energies, we show only energies with an absolute value smaller than  $ \frac{\Omega^2}{2\abs{\Gamma}} $, corresponding to frequencies inside the EIT transparency window. 

The number of bound states $N$ can be estimated \cite{Simon2005} by
\begin{equation}
	N \leq 1+2 \left\vert m\right\vert \int_{-\infty}^{\infty}\drm r\, \abs{r} W(r) .
	\label{eq:number_bound_states}
\end{equation}
This leads to a condition for the existence of only one bound state:
\begin{equation}
	\xi
	\leq
	\sqrt{3\sqrt{3}/\pi}
	\approx 1.2861
	\label{eq:single_bound_state}%
\end{equation}
Hence, a unique bound state exists only for small optical depth per blockade $\xi$. Consequently, to observe the formation of sufficiently long-lived photonic molecules in an experiment one hast to operate in this regime. Deeply bound states have small spatial extent, i.e., they are strongly localized and hard to excite by a flat initial photon distribution. The higher-$ n $ bound eigenstates $ \psi_n $, which exist for $ \xi\gg 1 $, are also hard to excite, since they exhibit many oscillations and, furthermore, are subject to strong decay, as we will show later-on. This explains the behavior seen in Fig.~\ref{fig:Figure1:c},\subref*{fig:Figure1:d}. The excitation of a bound photon state is only effective if a single bound state close to the continuum exists, i.e., in the weakly interacting limit. 
%
% =================================================
% =================================================

It is well known \cite{Abramov2001} that the bound state energies of a one-dimensional Schrödinger equation with a complex potential $W(r)$ are bounded by
\begin{equation}
	\left\vert E_{n}\right\vert \leq\frac{|m|}{4}
	\left(
	\int_{-\infty}^\infty\drm r\,
	\abs{W(r)}
	\right)^2.
	\label{eq:AbramovEnergyBound}%
\end{equation}
Making use of this inequality we then obtain the following estimate for the energy of the bound state
\begin{equation}
	|E_0|\leq\frac{1}{2}\,
	\xi^2
	\left(\frac{2\pi}{3}\right)^2
	\frac{2\Omega^2}{\Delta} \lesssim\frac{2\Omega^2}{\Delta},
	\label{BoundState_Energy estimation}
\end{equation}
where in the last step we assumed that only a single  bound state exists.
Hence the bound state energy is inside the low-frequency region of the EIT transparency window.

An estimate for the size of the bound state $\psi_{0}(r)  $ can be obtained from the uncertainty of the relative momentum. 
Assuming that we are in the regime of a single bound state close to the continuum, i.e., $ \xi\lesssim 1 $, a simple calculation shows that the momentum width of the bound state $\psi_0(r)$ is given by
\begin{equation*}
	\Delta p = 
	\int_{-\infty}^\infty\drm r\, \left(\frac{d\psi_0(r)}{dr}\right)^2 
	\lesssim
	\int_{-\infty}^\infty\drm r\, |2m W(r)|.
\end{equation*}
Using Heisenberg's uncertainty relation, we can derive an approximate expression for the size of the bound state
\begin{equation}
	r_\mathrm{b}
	=
	\int_{-\infty}^{\infty}\drm r\,r^2\psi_{0}^{2}(r) 
	\gtrsim
	\frac{1}{2}\left(\int_{-\infty}^{\infty}\drm r\,|2m W(r)|\right)^{-1},
%	\label{eq:size_bound_state}
\end{equation}
which yields
\begin{equation}
	r_\mathrm{b}\gtrsim\frac{3}{4\pi}\frac{\Labs }{\xi} > \Labs>\RB,
	\label{eq:boundStateSizeEstimation}
\end{equation}
where we used $ \xi\lesssim 1 $.
Thus in the parameter regime, where bound states can be excited, their spatial extend is rather large and exceeds the absorption length as well as the blockade radius. 
In Fig.~\ref{fig:singleBoundStateAndSpectrumHeff:BoundState} we show an eigenstate for $ \xi = 0.2 $ calculated by numerical diagonalization of the two-photon Hamiltonian compared to an exponential function with the size $ \frac{\pi}{3}\RB/\xi^2 $, showing a very good agreement.

% =================================================
\subsection{Bound-state components}
\begin{figure}
	\centering
	\subfloat[negative detuning, $ \Delta = -4\gamma $]{
%		\includegraphics[width = 0.5\columnwidth]{figs/Photon_Molecule_NegDelta}
%		\tikzset{external/force remake}
%		\setlength\fheight{0.6\columnwidth} 
%		\setlength\fwidth{0.8\columnwidth}	
%		\input{figs/Photon_Molecule_NegDelta.tikz} 	
		\includegraphics[]{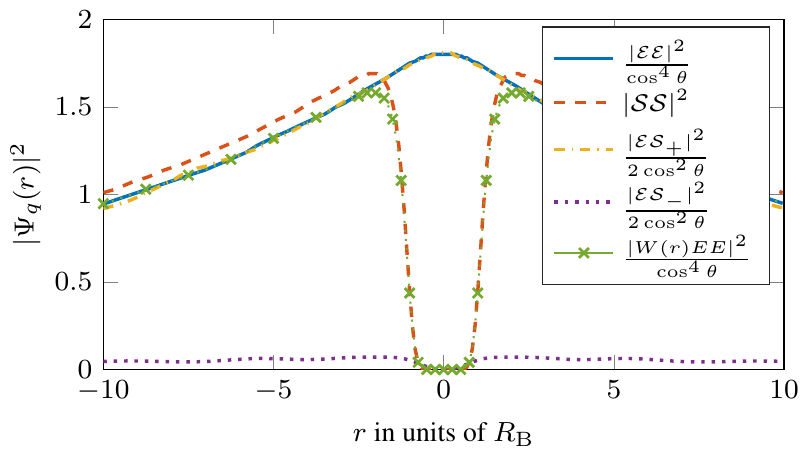}		
	}
	\\	
	\subfloat[positive detuning, $ \Delta = +4\gamma $]{
%		\tikzset{external/force remake}
%		\setlength\fheight{0.6\columnwidth} 
%		\setlength\fwidth{0.8\columnwidth}	
%		\input{figs/Photon_Molecule_PosDelta.tikz} 	
		\includegraphics[]{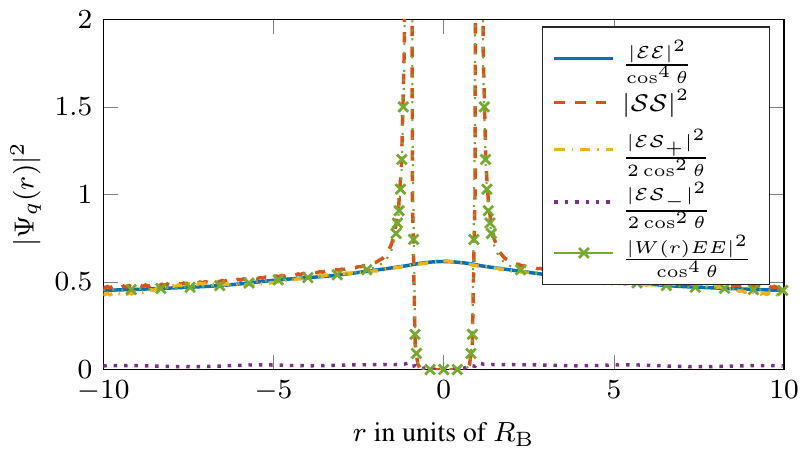}
	}
	\caption{%
	Photonic molecule state obtained from numerical time evolution of the paraxial Maxwell-Bloch equations for $K=0$, $ g=20\Omega $, $\xi = 0.2$, and $ t = 20 $ in units of $ \abs{\Delta}/2\Omega^2 $. Shown are the amplitudes of the wave-function components $ \EE $, $\mathcal{ES}_\pm$, $ \mathcal{SS} $, and $ W(r)\mathcal{EE} $, and each scaled with powers of $ \cos\theta $ according to Eq.~\eqref{eq:photon_molecule} to make them comparable. (a) shows the result for negative detuning and (b) shows the result for positive detuning, where the $ SS $-component exhibits resonances. Outside the blockade radius we find small deviations from the result we expect from Eq.~\eqref{eq:photon_molecule}.
}
\label{fig:photonic_molecule}
\end{figure}

% =================================================
%
The internal structure of the bound state can be found numerically by calculating the time evolution  of an initially broadly distributed wave function consisting of two dark-state polaritons for the case of vanishing center-of-mass momentum $ K=0 $. We assume that the $ \mathcal{SS} $-component has initially no excitation inside the blockaded region $ \abs{r}\lesssim\RB $. The results are shown in  Fig \ref{fig:photonic_molecule}. As expected, the $ \mathcal{SS} $-component is strongly suppressed inside the blockade radius. Here the bound state has mainly photonic character. In the case of $\Delta >0$ one recognizes a sharp peak of the $ \mathcal{SS} $-component close to the blockade radius. This coincides with the sharp minimum seen in the effective potential for positive detuning at this distance, cf.  Fig.~\ref{fig:effectivePotential:Positive}.

If condition \eqref{eq:single_bound_state} for a single bound state is fulfilled, the amplitude $ \EE(r,t) $ of finding two photons at relative distance $ r $ 
reads
\begin{multline}
	\frac{\EE(r,t)}{\cos^{4}\theta}
	=
	C_{0} \mathrm{e}^{-\I E_{0}t}  \psi_{0}(r)  
	+\int C(E)\mathrm{e}^{-\I Et}\psi_E(r)\drm E, 
	\label{eq:photon_wavepaket}
\end{multline}
where  $C_{0}$ and $C(E)$ are the overlap integrals between the initial state and the bound and continuum eigenstates, respectively. We here consider only the first part, corresponding to the bound state and will discuss the continuum states in the following section.The amplitudes of the remaining components $\mathcal{ES}, \mathcal{SE}$ and $\mathcal{SS} $, of the bound state can be obtained by substituting the solution for $\EE$ into the two-particle Schrödinger equation. A direct calculation gives 
\begin{equation}
	\mathcal{ES}_+(r,t) = \mathcal{ES}+\mathcal{SE}
	\approx
	-2C_{0}\cos^{3}\theta\psi_{0}(r) \mathrm{e}^{-\I E_0t}.
	\label{Sym_Spin_Photon}
\end{equation}
In obtaining this expression we have assumed that $g^2/|\Delta|\gg cK,|E_{0}|$. 
The calculation of the spin component $\mathcal{SS}(r,t) $ is more involved, but straightforward. After simple algebra we arrive at
\begin{equation}
	\mathcal{SS}(r,t)
	\approx C_{0}\frac{\cos^{2}\theta}{1-\frac{\Delta}{2\Omega^{2}}V(r)}\psi_{0}(r)
	 \mathrm{e}^{-\I E_0t}.
	 \label{SpinSpin_Component}
\end{equation}
Analogous calculations can be performed for the antisymmetric component $ \mathcal{ES}_-=\mathcal{ES}-\mathcal{SE} $, 
which becomes negligible if the size of the bound state is much larger than the off-resonant optical length, $r_\mathrm{b}\gg\Labs$. 
\begin{equation}
	\frac{\abs{\mathcal{ES}_-(r,t)}}
	{2C_{0}\cos^{3}\theta\psi_{0}(r)}\approx\left\vert \frac{\Labs }{\psi_{0}( r)}
	\frac{d\psi_{0}(r)  }{dr}\right\vert \rightarrow 0
\label{Antysymmetric}%
\end{equation}
In this weak-interaction limit the amplitudes, Eqs.~\eqref{eq:photon_wavepaket} --\eqref{SpinSpin_Component} and \eqref{Antysymmetric}, can be combined in a compact form, $\vec{\Psi}=(\mathcal{EE}, \mathcal{ES}, \mathcal{SE}, \mathcal{SS})^{T}$, yielding
\begin{equation}
	\bm{\Psi}(r,t)=\cos^2\theta C_{0}
	\begin{pmatrix}
	\cos^{2}\theta\\
	-\cos\theta\\
	-\cos\theta\\
	\frac{1}{1-\frac{\Delta}{2\Omega^{2}}V(r)}
	\end{pmatrix}
	\psi_{0}(r)  \mathrm{e}^{-\I E_{0}t},
	\label{eq:photon_molecule}
\end{equation}
where the factor $ \cos^2\theta $ in front appears as a result of projecting the initial state onto the state of two free polaritons and can be changed by choosing a specific initial state vector $\vec\Psi(r,0)$. 
Note that this result is only applicable, when  the energy $\abs{E_0}$ of the bound state is much smaller than other energies involved in the system, e.g. $2\Omega^{2}/\abs{\Delta}$. The spatial size of the bound state $\psi_{0}( r)  $ is in this case larger than $\Labs$. 
We observe that $ \vec\Psi(r,t) $ in equation \eqref{eq:photon_molecule} describes a two-photon wave packet that, although subject to decay, propagates form-stable and exhibits bunching for small distances, i.e., a \emph{photonic molecule} state.

% =================================================
% =================================================

We find relatively good agreement of the different asymptotic forms of the wave function amplitudes, Eq.~\eqref{eq:photon_molecule}, with the numerical results. There is only a small deviation, as we still find a finite remaining antisymmetric component $ \mathcal{ES}_{-}(r,t) $ and correspondingly a slightly increased spin excitation.

% =================================================
\section{Bound states and continuum}
% =================================================
\label{sec:boundStatesAndContinuumStates}
A major goal of this paper is to analyze the interplay between bound state and continuum contributions. In this section we employ an effective Schrödinger equation to derive asymptotic analytical solutions of Eq.~\eqref{eq:GreenFunctionIntegralEquationSmallFreq} including both, bound and continuum parts, that we compare to numerical solutions. We show that an initial state evolves into a superposition of bound states and scattering states both of which contribute to the bunching signal.

%
% ================
\subsection{Approximate analytic solutions}
% ================
%
For the relevant propagation distances the dispersive nature of the interaction potential, i.e., the frequency-dependence of $ W(r,\omega) $, can be ignored for the dynamics of the two-particle wave function (see appendix \ref{sec:appendix_a}).
In this case we can proceed in a standard way and reformulate the integral equation, Eq.~\eqref{eq:GreenFunctionIntegralEquationSmallFreq}, as a Schrödinger-type initial value problem for the propagator $G(r,r',t)$,
\begin{equation}
	\begin{gathered}
		\I\frac{\partial}{\partial t} G(r,r',t)   
		= \left(\frac{1}{2m}\frac{\partial^2}{\partial r^2}+W(r)\right)G(r,r',t),
		\\
		G(r,r',0)   
		= \delta(r-r'). 
	\end{gathered}	
	\label{eq:Propagator_Interaction}
\end{equation}
In obtaining this result we have omitted the kinematic term $cK\cos^{2}\theta$ for the center-of-mass motion, as it generates a trivial shift in time. In the following we thus assume $K=0$. The potential $W(r)$, \eqref{eq:effective_potential}, can be treated as an effective interaction between two photons. 
 At short times the model \eqref{eq:Propagator_Interaction} does not approximate the full dynamics well, since our derivations of Eq.~\eqref{eq:Propagator_Interaction} are based on the assumption that the evolution time should be long compared to all other characteristic time scales of the system (see: appendix \ref{sec:appendix_a}).

% ================
\begin{figure}
	\centering
%		\tikzset{external/force remake}
%		\setlength\fheight{0.6\columnwidth} 
%		\setlength\fwidth{0.8\columnwidth}	
%		\input{figs/EE_FunctionOfRelativeDistance.tikz} 
		\includegraphics[]{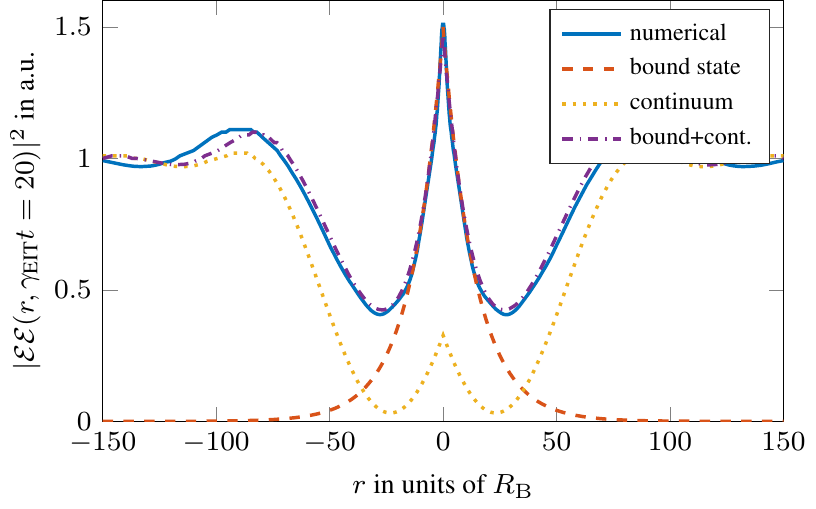}
		\caption{%
		Second order correlation functions $ |\EE(r,t)|^2 $ of two photons as function of relative distance $ r $ and fixed time $t = 20$ (in units of $ \abs{\Delta}/2\Omega^2$). The solid blue line shows a numerical calculation for $ K=0 $, $ g/\Omega = 100 $, $ \Delta = -4\gamma $ in the weakly interacting regime with $ \xi = 0.2 $. The dashed red and the dotted yellow line show the bound and continuum part of the wave function, respectively, according to Eq.~\eqref{eq:photonicSolutionDeltaFunction}, and the dash-dotted purple line shows the sum of both.
		}
		\label{fig:Compare_Model}
\end{figure}
% ================

As can be seen from the full numerical solution in Fig.~\ref{fig:Compare_Model}, for weakly interacting photons, i.e., $\xi\ll 1$, the range of spatial variation of the two-photon amplitude $\EE(r,t)$ is much greater than the range of the potential, i.e., the blockade radius $\RB$ . This suggests that $W(r)$ can be approximated by a delta-like pseudopotential.
\begin{equation}
	W(r)\rightarrow W_\text{eff}(r)=
	\frac{2\pi \RB }{3}\frac{2\Omega^2}{|\Delta|}
	\frac{1}{(1+\I{\gamma}/{|\Delta|})^{5/6}}\delta(r).
	\label{Delta_Potential_Approximation}
\end{equation}
Assuming that the initial two-photon amplitude is uniformly distributed in the relative coordinate $ r $ one can show that the Schrödinger equation for $\EE(r,t)$
with the effective interaction potential $W_\text{eff}(r)$
 admits analytical solutions in closed form  at large times. For convenience we introduce dimensionless time and space coordinates that are measured in units of $(2\Omega^{2}/\abs{\Delta})^{-1}$ and $\RB$, respectively. After a lengthy but straightforward calculation we find the following expression for the two-photon amplitude $\EE(r,t) $,
\begin{multline}
	\frac{\EE(r,t)}{\cos^4\theta}
	=
	\erf\Bigl(\sqrt{\tfrac{i\beta}{2t}}\abs{r}\Bigr)	
	+\exp\Bigl(-\I\frac{\beta\eta^{2}}{2}t-\beta\eta\abs{r}\Bigr)\\
	\times
	\left\{2-\Bigl[1+\erf\Bigl(-\sign[\Re(\beta\eta)]
	\sqrt{\tfrac{\beta\eta
^{2}}{2\I}t}+\sqrt{\tfrac{\I\beta}{2 t}}\abs{r} \Bigr) \Bigr] \right\},
\label{eq:photonicSolutionDeltaFunction}
\end{multline}
where for convenience we defined the constants
\begin{equation}
	\eta
	=
	\frac{2\pi}{3}\frac{1}{(1+\I\frac{\gamma}{|\Delta|})^{5/6}},
%	\label{Amplitude_Interaction}
	\qquad
	\beta 
	=
	\frac{1}{2}\frac{\xi^2}{1+\I\frac{\gamma}{|\Delta|}}.
	\label{Mass_New_Units}
\end{equation}
The term $2\exp\bigl(-\I\frac{\beta\eta^{2}}{2}t-\beta\eta\abs{r}\bigr)$ in Eq.~\eqref{eq:photonicSolutionDeltaFunction} corresponds to a single bound state wavefunction of the effective potential $W_\text{eff}(r)$, if the condition $ \Re(\beta\eta)>0 $ is fulfilled. This holds, if $ \abs{\Delta}>0.8665\,\gamma $,
i.e., under off-resonant driving conditions.
The size of the bound state (in units of $\RB$) is equal to
\begin{equation}
	r_\mathrm{b} \approx (\beta\eta)^{-1}
	\approx\frac{\pi}{3}
	\xi^{-2}
	\gg 1.
%	\label{BoundStates_Size}
\end{equation}

In Fig.~\ref{fig:Compare_Model} we show the bound and continuum-state contributions obtained from Eq.~\eqref{eq:photonicSolutionDeltaFunction} and compare them to the full
numerical solution. One recognizes very good agreement, which also shows that 
the approximation used to derive Eq.~\eqref{eq:Propagator_Interaction} is justified. One notices that the spatial structure of bound and continuum states near $r=0$ is the same.

The complex energy of the bound states can be read off from Eq.~\eqref{eq:photonicSolutionDeltaFunction}. Up to second order in $ \gamma/\Delta $ it is given by
\begin{align}
	E_0
	&\approx
	-\frac{\pi^2}{9}\xi^2\left(1-\I\frac{8}{3}\frac{\gamma}{\Delta}-\frac{44}{9}\frac{\gamma^2}{\Delta^2}\right)
\end{align}
From this we can also read off the decay rate of the bound state which is approximately
\begin{equation}\label{eq:boundStateDecayRate}
\gamma_\mathrm{b}\approx 2.924\, \xi^2 \frac{\gamma}{\Delta}.
\end{equation}
Note that both $E_0$ and $\gamma_\mathrm{b}$ are in units of $2\Omega^2/\vert \Delta\vert$.
One recognizes that long lifetimes of bound states require small optical depth per blockade volume, $\xi$.

% ================
\subsection{Bound-state and continuum contributions to bunching}
% ================

\begin{figure}
	\centering
	\centering
	\subfloat[Small single photon detuning $ \Delta = -1.5\gamma $]{
%		\tikzset{external/force remake}
%		\setlength\fheight{0.5\columnwidth} 
%		\setlength\fwidth{0.75\columnwidth}	
%		\input{figs/amplitudeAsFunctionOfTimeSmallDelta.tikz} 
		\includegraphics[]{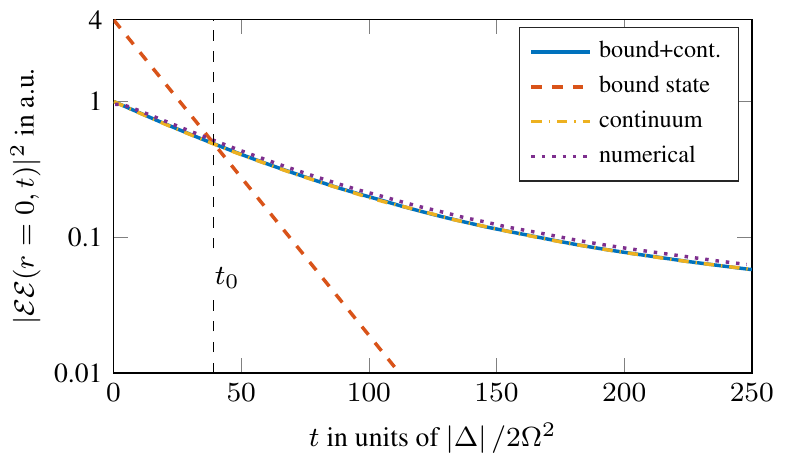}
		\label{fig:Amplitudes:SmallDelta}
		}\\
	\subfloat[Large single photon detuning $ \Delta = -12\gamma $]{
%		\tikzset{external/force remake}
%		\setlength\fheight{0.5\columnwidth} 
%		\setlength\fwidth{0.75\columnwidth}	
%		\input{figs/amplitudeAsFunctionOfTimeLargeDelta.tikz} 
		\includegraphics[]{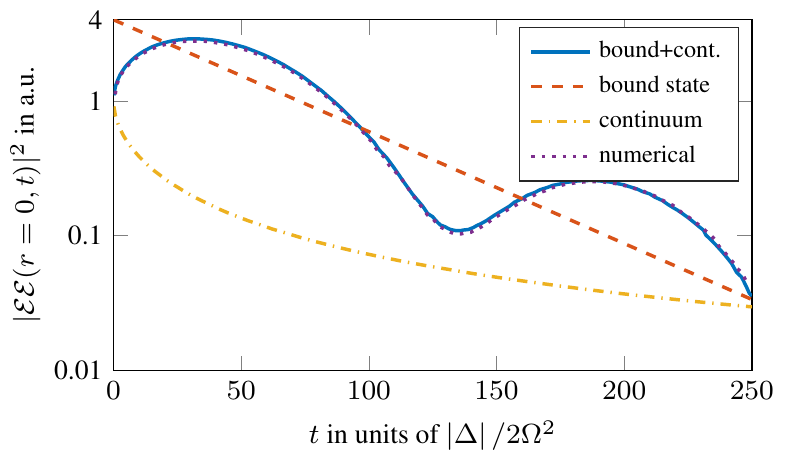}
		\label{fig:Amplitudes:LargeDelta}
	}	
	\caption{%
		Logarithmic plot of the amplitudes of bound- (dashed, red) and continuum-states components (dashed-dotted, yellow) of the two-photon wave function $\mathcal{EE}(0,t)$ as function of time for zero relative distance shown (solid, blue). The dotted purple line shows the full numerical solution. The results are in the weakly interacting regime for $ \xi = 0.2 $ and $ g/\Omega =100 $ and calculated for (a) small single photon detuning $ \abs{\Delta } = 1.5\gamma $ and (b) large detuning $ \abs{\Delta}= 12\gamma $. The dashed vertical line in (a) indicates the crossover time scale $ t_0 $.}
	\label{fig:Amplitudes}
\end{figure}

There are two distinct features of bound and continuum states. First of all, while in the vicinity of $r=0$ the continuum states have the same spatial structure as the bound state, they are the dominant contribution at large relative distances $r$,
see Fig.\ref{fig:Compare_Model}. This is due to the exponential localization of the bound state on a length scale $r_\mathrm{b}$.
Secondly, as can be seen from Eq.~\eqref{eq:photonicSolutionDeltaFunction},
bound and continuum contributions have a different time evolution. While the continuum states decay diffusively in time, i.e., $\propto 1/\sqrt{t}$, bound states decay exponentially. This is illustrated in Fig.~\ref{fig:Amplitudes}, where we have plotted the amplitudes of bound and continuum state as function of time 
at vanishing relative distance $r=0$ along with the two-photon amplitude $\mathcal{EE}(0,t)$. The larger the detuning $\vert \Delta\vert$ the slower the decay of the bound state.
Nevertheless for large times the continuum contributions become the dominant part also for small relative distances. The oscillatory behavior of $\mathcal{EE}$ is an interference
effect between bound and continuum contributions, which will be discussed in more detail later.

% ================
% ================

At large times, the solution of \eqref{eq:photonicSolutionDeltaFunction} at $r=0$, can be further simplified to
\begin{equation}
	\frac{\mathcal{EE}(0,t)  }{\cos^{4}\theta}\approx2\exp\Bigl(-
	\frac{\I\beta\eta^{2}}{2}t\Bigr)  -\frac{1}{\sqrt{\frac{\pi\beta\eta^{2}}{2\I}t}},
	\label{eq:Simplified_Delta}
\end{equation}
where again the first term on the right hand side corresponds to the bound state while the second term gives the contribution from the continuum.

Using Eq.~\eqref{eq:photonicSolutionDeltaFunction} or, for sufficiently large $ \abs{\Delta} $, Eq.~\eqref{eq:Simplified_Delta} we can identify a crossover time $t_0$ at which the contribution of the scattering states becomes the dominant one. This also means that for $t \gg t_0$ any observed bunching is solely due to the scattering states. A simple analysis shows that  $t_0$ is minimal if 
\begin{equation}
	\Re(\beta\eta^{2}) = 0,
	\label{half_resonance}
\end{equation}
i.e., when $\gamma/\left\vert \Delta\right\vert =\tan\frac{3\pi}{16}\approx0.6681\approx 2/3$ 
and at this point for $t_{0}$ one has
\begin{equation}
	t_{0}\approx{} \frac{\pi}{2\xi^2},
	\label{eq:critical_time}
\end{equation}
according to Eq.~\eqref{eq:photonicSolutionDeltaFunction}. For $ \xi<1 $ this time scale is much larger than one in units of the typical EIT time scale $ \abs{\Delta}/2\Omega^2$.
%

% ================
\section{Filtering of bound and continuum components}
% ================
As can be seen in Fig.~\ref{fig:Amplitudes} the two-photon amplitude shows an oscillatory time dependence. These oscillations result from an interference between bound- and continuum-states contributions to the two-photon amplitude, due to the different phases of these terms. We will now investigate the time evolution of the phase in more detail and will argue that this can be used to filter out the bound-state components, allowing for an experimental investigation of the photonic molecules alone.

\begin{figure}
	\centering
	\subfloat[$|\EE(r,t)|^2$ in a.u.]{
%	\includegraphics[width=0.5\columnwidth]{figs/timeEvolutionRelativeCoordinateDelta12}}
%		\tikzset{external/force remake}
%		\setlength\fheight{0.25\columnwidth} 
%		\setlength\fwidth{0.45\columnwidth}	
%		\input{fig7Amplitude.tikz} 
		\includegraphics[]{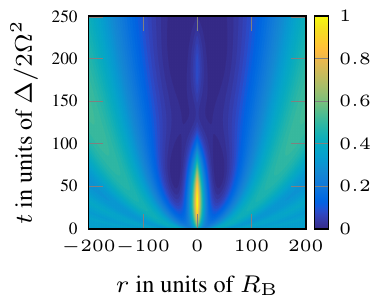}
	}
	\subfloat[$\arg(\EE(r,t))$ in units of $ \pi $]{
%		\includegraphics[width=0.5\columnwidth]{figs/Phase_xi0-2_Delta12}
%		\tikzset{external/force remake}
%		\setlength\fheight{0.25\columnwidth} 
%		\setlength\fwidth{0.45\columnwidth}	
%		\input{fig7Phase.tikz}	
		\includegraphics[]{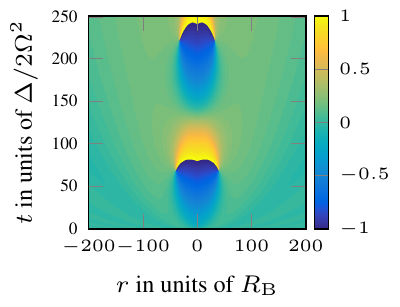}
	}
	\caption{Time evolution of two excitation wave function for $ K=0 $ inside medium in the weakly interacting regime where $ \xi= 0.2 $ for $ \Delta = -12\gamma $, and $ g=100\Omega $. (a) Amplitude $|\mathcal{EE}(r,t)|$ and (b) phase $\arg(\EE(r,t))$
	 One recognizes a large phase, constant over the extend of the two-photon bound state for fixed times.
	} % 
	\label{fig:t_evol_inside}
\end{figure}

In order to employ the phase shift of the photonic molecule for its experimental separation, it should be spatially homogeneous and at the same time sufficiently large. In Fig. \ref{fig:t_evol_inside} we show amplitude and phase of $\mathcal{EE}(r,t)$ as functions of relative distance $r$ and time $t$ obtained from numerically solving the full two-particle evolution. 
One recognizes that the phase shift is large and constant in space over the whole extend of the localized two-photon component for fixed times. 
% ================
% ================
This is fully consistent with the approximate analytic solutions obtained in Sec.\ref{sec:boundStatesAndContinuumStates}. 

As can be seen from Eq.~\eqref{eq:Simplified_Delta}, the bound state attains a dynamical phase
\begin{equation}
\phi_\mathrm{b}(t) = \tfrac{1}{2}\Re(\beta\eta^2)t
\label{eq:phi_b}
\end{equation}
At the point $\gamma/\left\vert \Delta\right\vert =\tan\frac{3\pi}{16}$, where $t_{0}$ is minimal, this
dynamical phase vanishes, as $\Re(\beta \eta^2)=0$.

\begin{figure}
	\centering
	\centering
	\subfloat[Small single photon detuning $ \Delta \approx -1.5\gamma $]{
%		\tikzset{external/force remake}
%		\setlength\fheight{0.5\columnwidth} 
%		\setlength\fwidth{0.75\columnwidth}	
%		\input{figs/phasesAsFunctionOfTimeSmallDelta.tikz}
		\includegraphics[]{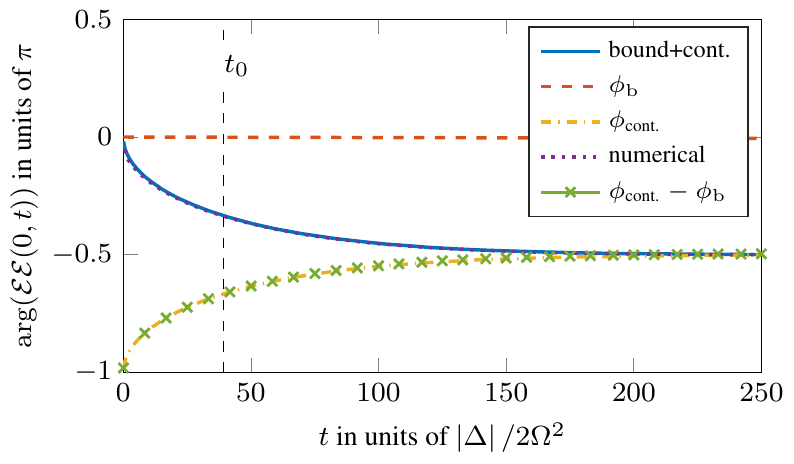}
		\label{fig:Phases:SmallDelta}
	}\\
	\subfloat[Large single photon detuning $ \Delta = -12\gamma $]{
%		\tikzset{external/force remake}
%		\setlength\fheight{0.5\columnwidth} 
%		\setlength\fwidth{0.75\columnwidth}	
%		\input{figs/phasesAsFunctionOfTimeLargeDelta.tikz} 
		\includegraphics[]{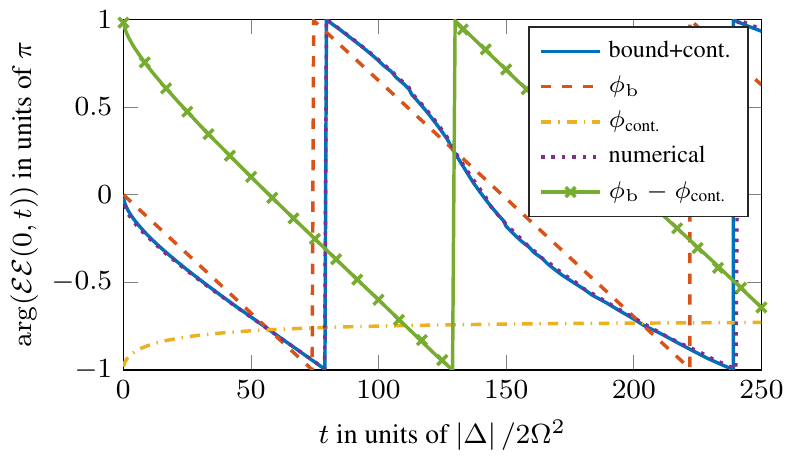}
		\label{fig:Phases:LargeDelta}
	}	
	\caption{%
		Phases $ \phi_\textrm b $ and $ \phi_\text{cont.} $ of bound-state (dashed, red) and  continuum-state components (dashed-dotted, yellow) of the two-photon wave function $\mathcal{EE}(0,t)$ (solid, blue) , respectively, as function of time for zero relative distance shown. The dotted purple line shows the phase of a full numerical solution. The green crosses (green line with crosses) shows the phase difference between bound and continuum states.	The results are in the weakly interacting regime for $ \xi = 0.2 $ and $ g/\Omega =100 $ and calculated for (a) the special case $ \abs{\Delta } \approx 1.5\gamma $, where $t_0$ is minimal and the continuum reaches a phase of $ \pi/2 $ (b) large detuning $ \abs{\Delta}= 12\gamma $.}
	\label{fig:Phases}
\end{figure}

At the same time, the phase of the continuum states approaches for large times the constant value 
\begin{equation}
\phi_\text{cont} (t) = -\pi/2,
\label{eq:phi_c}
\end{equation}
for $\gamma/\left\vert \Delta\right\vert =\tan\frac{3\pi}{16}$. Going to larger detunings the bound state attains a nonvanishing dynamical phase and
the phase of the continuum states increases with $ \abs{\Delta} $ up to a value of
\begin{equation}
\phi_\text{cont} (t) =  -3\pi/4,
\label{eq:phi_c-2}
\end{equation}
for very large $ \abs{\Delta} $. This is further illustrated in Fig.~\ref{fig:Phases}, where we have plotted the phases of the bound and continuum states as function of time for the detuning of $\Delta \approx -\tan\frac{3\pi}{16}\gamma$ and a larger detuning of $ \Delta = -12\gamma $. We have verified the accuracy of the approximate solutions by comparison to numerical solutions of the Maxwell-Bloch equations in the case $ K=0 $.
% ================
% ================

We note that this phase of the continuum state is very robust, as it only depends on $ \abs{\Delta}/\gamma $ and can be tuned by changing the frequencies of the probe and control fields.

When combining our results about the amplitudes and phases of the bound state and continuum state components one can distinguish three regimes depending on the ratio
$ \abs{\Delta}/\gamma $. First, for small detuning the continuum states dominate the dynamics at all times and attain a phase of $ -\pi/2 $. Secondly, for intermediate detuning the crossover time $ t_0 $ increases and the dynamics is governed by an interplay of bound and continuum state components, and finally, for large detuning the continuum state decays very quickly and the bunching is solely due to the bound state. To observe the photonic molecule, one could simply go to large detuning and wait until the continuum states are decayed. However, this would be experimentally challenging. Therefore it is better to work in the regime of intermediate detunings. As can be seen from Fig.~\ref{fig:Phases}, in this case the phase of the continuum contribution attains its robust asymptotic value long before the crossover point $ t_0 $. Thus, the continuum contribution can effectively be filtered out by interferometric techniques, as sketched in Fig.~\ref{fig:Figure1:b}, allowing for an isolation and observation of the probe field component corresponding to the molecular state.

% ================
\section{Conclusion}
% ================

We discussed the bunching of dark-state polaritons propagating under conditions of electromagnetically induced transparency in a gas of atomic three-level atoms and interacting via van der Waals-type interactions mediated by Rydberg interactions of the atoms. By employing a Green's function approach, we derived an effective model for two dark-state-polariton excitations and analyzed its spectral properties, showing the existence of bound eigenstates. We showed that for weak interactions, quantified by the optical depth per blockade, and in an off-resonant driving scheme the model has a single eigenstate close to the scattering continuum. We argued that, while the higher-$ n $ bound states are difficult to excite, this low-energy single bound state can experimentally observed. We confirmed this by numerical integration of the full Maxwell-Bloch equations for two particles which shows bunching for sufficiently small values of the optical depth per blockade, but anti-bunching for larger values, as has also be shown in recent experiments \cite{Peyronel2012a,Firstenberg2013}. 

By using the Green's function approach we showed that this bunching feature cannot solely explained by the bound eigenstate, but rather comes about by an interplay of bound and continuum states. We derived closed analytic expressions for the bound state and continuum wave functions in the limit of weak interactions, where the effective interaction potential can be approximated by a $ \delta $-potential. This expressions allowed us to investigate the time-dependence of the individual components. Specifically, we showed that the bound state decays exponentially in time, whereas the scattering states have an diffusive time-dependence. Thus, for small times the bunching has to be explained by a superposition of bound and continuum wave function, while for large times the polariton pair is dominated by the continuum. Moreover, we found that, after some time, the continuum component attains a robust and constant phase, while the bound state exhibits a dynamical phase. This allows to filter bound and continuum components by making use of a homodyne detection scheme. We here concentrated on an effective one-dimensional setting. In three spatial dimensions there is an additional constraint for the existence of a bound state, which we discuss in appendix~\ref{sec:app:bound_states_in_3D}.

% =============== bibliography ==================
\bibliographystyle{apsrev4-1}
\bibliography{photonMolecules.bib}
% ===============================================

\appendix
% ================================
\section{Green's function approach}\label{sec:appendix_a}
% ================================
Here we present the details of the Green's function approach employed in the derivation of Eqs.~\eqref{eq:photonicEvolutionAsymptotic} to \eqref{eq:effective_potential}. The noninteracting result, Eq.~\eqref{eq:freeGreensFunction} follows then immediately from Eq.~\eqref{eq:photonicEvolutionAsymptotic} when setting the effective potential $W$ equal to zero.

To simplify the derivation in the following we consider an initial vector $\ket{\Psi_0}\equiv\vec{\Psi}(K,r,t=0)$ given by
\begin{gather}
	\vec{\Psi}( K,r,t=0)  =f(  K,r)  \left\vert
	\varphi_1\right\rangle, 
	\label{app:eq:initialVector}
\end{gather}
where $\ket{\varphi_1}=(1,0,0,0)^{T} $ denotes a pure field excitation, i.e., $\mathcal{EE}$. The derivation with other initial conditions can be treated analogously. 
The time evolution of the two-photon amplitude $\mathcal{EE}(t)$ for  admits a spectral Fourier-Laplace representation, given by
\begin{equation}
	\EE(t) = \frac{1}{2\pi \I}%
	\int\limits_{-\infty}^{\infty}
	\mathrm{e}^{-\I\omega t}
	\bra{\varphi_1}
	\hat G(\omega)
	\ket{\Psi_0}\drm\omega,\ t>0,
	\label{A3}%
\end{equation}
for $t>0$. The full and  free Green's functions are defined by
\begin{equation}\label{app:eq:GreensFunctions}
	\hat G(\omega)\equiv \frac{1}{\Hcal-\omega-\I0^+}
	\text{\ \ and\ \ }
	\hat G_0(\omega)\equiv \frac{1}{\Hcal_0-\omega},
\end{equation}
respectively, where $	\Hcal=\Hcal_{0}+V(r)\hat{\mathrm{P}}_{\mathcal{SS}}$ as in the main text. We denote the Green's function governing the time evolution of $ \mathcal{EE}(t) $ by
\begin{equation}
	G_{11}(\omega)  = \bra{\varphi_1}\hat G(\omega)\ket{\varphi_1}.
	\label{app:eq:G11}
\end{equation}
This operator usually has a branch cut and in the presence of an interaction
between atoms it may have poles, which correspond to resonant states with
negative imaginary parts ($t>0$). For sufficiently large times, larger than
the decay time of resonances, branch cut singularities of the Green's function
contribute into the integral \eqref{A3} only.

The Green's function \eqref{app:eq:G11} satisfies the operator equation%
\begin{multline}
	G_{11}(\omega)  =
	\bra{\varphi_1}\hat G_0(\omega)\ket{\varphi_1}\\
	-\bra{\varphi_1}\hat G_0(\omega)\ket{\varphi_4}V\bra{\varphi_4}\hat G(\omega)\ket{\varphi_1},
	\label{A5}
\end{multline}
where the propagators in this equation can be written in the form
\begin{gather}
	\bra{\varphi_1}\hat G_{0}(\omega)\ket{\varphi_1}
	=
	\alpha_{11}\left(	\omega\right)  +\gamma^{2}(\omega)  g_{00}(\omega)
	, \label{A5a}\\
	\bra{\varphi_1} \hat G_{0}(\omega)  \ket{\varphi_4} 
	=
	\gamma(\omega)g_{00}(\omega),
\end{gather}
with the quantities
\begin{align}	
	\alpha_{11}(\omega) 
	&  =\frac{\I\Gamma}{\left(
	cK-\omega\right)  \I\Gamma-2g^{2}},\\
	\gamma(\omega)
	&=\frac{2\Omega^{2}-\I\omega\Gamma}{2g^{2}-\I(\omega-cK)  \Gamma
	}\\
	g_{00}(\omega)
	&=\frac{1}{\frac{p^{2}}{2m_{0}(\omega)  }-\Lambda_{0}(\omega)  },
\end{align}
the effective mass
\begin{equation}
	m_{0}(\omega)   
	=\frac{\Omega^{2}g^{2}\left(  g^{2}+\Omega
	^{2}+\left(  \frac{cK}{2}-\omega\right)  \I\Gamma\right)  }{\I\Gamma\left(
	2i\Omega^{2}+\omega\Gamma\right)  ^{2}c^{2}}\\
\end{equation}
and
\begin{multline}
	\Lambda_{0}(\omega)   
	=(2\Omega^{2}-\I\omega\Gamma)(2 g^2+2\Omega^2+\I cK\Gamma-2\I\omega\Gamma)\\
	\times\frac{
	(2\omega g^{2}-(cK-\omega)(2\Omega^{2}%
	-\I\omega\Gamma))}
	{4g^{2}\Omega^{2}(2 g^2-\I\Gamma(\omega-cK))  }.
\end{multline}

In the next step we find the equation for the Green's function $\bra{\varphi_4}\hat G(\omega)\ket{\varphi_1}$.
It can be derived in an analog manner to Eq.~\eqref{A5} which yields
\begin{multline}
	( 1+\alpha_{00}(\omega)  V)  \bra{\varphi_4}\hat G(\omega)  \ket{\varphi_1}\\
	=
	\bra{\varphi_4} \hat G_0(\omega)  \ket{\varphi_1}-g_{00}(\omega)  V\bra{\varphi_4} \hat G(\omega)  \ket{\varphi_1} ,  \label{A7}%
\end{multline}
where $\alpha_{00}(\omega) = \frac{\I\Gamma}{2\Omega^{2}-\I\omega\Gamma}$. Absorbing the factor $ (1+\alpha_{00}(\omega)V) $ into the Green's function $ G_{41} $ by defining
\begin{equation}
	G_{41}(\omega)  
	=
	(1+\alpha_{00}(\omega)V)
	\bra{\varphi_4} G(\omega) \ket{\varphi_1} ,  
	\label{A9}
\end{equation}
we can write the operator equation \eqref{A7} in the closed form
\begin{equation}
	G_{41}(\omega)  
	=
	\gamma(\omega)g_{00}(\omega)
	- g_{00}(\omega)  W(\omega)	G_{41}(\omega),  
	\label{A10}
\end{equation}
where the effective potential $ W(r,\omega) $ is defined by
\begin{equation}
	W(r,\omega)  
	=
	\frac{V(r)}{1+\alpha_{00}(\omega)  V(r)}.
	\label{eq:app:effectivePotentialFreqDep}
\end{equation}
Combining Eq.~\eqref{A10} with Eq.~\eqref{A5}, we arrive at the following equation
\begin{equation}
	G_{11}(\omega)  
	=
	\alpha_{11}(\omega)+\gamma(\omega)G_{41}(\omega)   
	\label{A12}
\end{equation}
for the required Green's function $G_{11}(\omega)  $. The
evolution of the two-photon amplitude, in the coordinate representation,
\begin{multline}
	\EE(r,t)   
	= 
	\frac{1}{2\pi \I}\int_{-\infty}^{\infty}
	\mathrm{e}^{-\I\omega t}\alpha_{11}(\omega)\drm \omega
	\\
	+
	\frac{1}{2\pi \I}\iint_{-\infty}^{\infty}
	\mathrm{e}^{-\I\omega t}  \gamma(\omega)  G_{41}(r,r',\omega)f(K,r') \drm\omega \drm r'
\end{multline}
is obtained by substituting Eq.~\eqref{A12} into the integral \eqref{A3}.

We note that the first integral at large $t\gg \frac{\Delta^{2}}{2\gamma g^{2}}$
is negligible. We thus have
\begin{equation}
	\mathcal{EE}(r,t) =
	\frac{1}{2\pi \I}
	\iint
	\mathrm{e}^{-\I\omega t}  \gamma^{2}(\omega)  G\left(
	r,r' ,\omega\right)  f\left(  K,r' \right)  \drm\omega \drm r',
	\label{A13}
\end{equation}
where
$ G(r,r' ,\omega) = G_{41}(\omega)/\gamma(\omega)$,
which obeys the following integral equation,
\begin{multline}
	G\left(  r,r' ,\omega\right) = g_{00}(r,r',\omega) \\
	-  \int\limits_{-\infty}^{\infty}
	g_{00}\left(  r,r^{\prime\prime},\omega\right)  W\left(  r^{\prime\prime
	},\omega\right)  G\left(  r^{\prime\prime},r' ,\omega\right)
	\drm r^{\prime\prime}.\label{A14}%
\end{multline}
As was pointed out earlier, our interests are restricted to large times.
Thus we can further simplify the expressions for the Green's functions by considering only small frequencies and momentum in Fourier space. In particular, we assume low frequencies
$ \omega\ll \min (2\Omega^{2}/\abs{\Gamma},2g^{2}/\abs{\Gamma})  $ 
and small center-of-mass momentum 
$ cK\ll \min (2\Omega^{2}/\abs{\Gamma},2g^{2}/\abs{\Gamma})  $.
In this limit all quantities such as 
$m_{0}(\omega)  ,\Lambda_{0}(\omega)  $ and $\gamma(\omega)  $ take much simpler forms, and, moreover, the effective potential, Eq.~\eqref{eq:app:effectivePotentialFreqDep} becomes independent of $ \omega $. This leads to the following expression for $\mathcal{EE}(r,t)$,
\begin{equation}
	\EE(r,t)
	=
	\frac{\cos^{4}\theta}{2\pi \I}\iint_{-\infty}^{\infty}	\mathrm{e}^{-\I\omega t}  G(r,r',\omega)  f(K,r')  \drm\omega \drm r', 		
\label{A15}
\end{equation}
where the Green's function $G(r,r',\omega)$ satisfies the following integral equation
\begin{multline}
	G(r,r' ,\omega) = G_{0}(r,r',\omega)\\
	- \sin^{4}\theta
	\int_{-\infty}^{\infty}
	G_{0}(r,r^{\prime\prime},,\omega)  W\left(  r^{\prime\prime}\right)
	G\left(  r^{\prime\prime},r' ,\omega\right)  \drm r''.
	 \label{A16}%
\end{multline}
The free Green's function
\begin{equation}
	G_{0}(\omega) = \frac{1}{\frac{p^{2}}{2m}-\omega
	+ cK\cos^{2}\theta}, \label{A17}%
\end{equation}
describes a Schrödinger particle with the complex effective mass
\begin{equation}
	m 
	= 
	\I \frac{g^2\OmegaEff^{2}}{4\Gamma\Omega^{2}c^{2}}
	=
	\I\frac{g^2}{4c\Gamma\vgroup}.
	\label{app:eq:reduced_mass}
\end{equation}

Note that the solution of the Schrödinger problem, in integral representation, Eq.~\eqref{A15}, will be of little help for any practical purposes.
However, the Eqs.~\eqref{A15} to \eqref{A17} allow to show that
the time evolution of the two-photon amplitude $\mathcal{EE}\left(
r,t\right)  $ obeys the following Schr\"{o}dinger equation
\begin{equation}
\I\frac{\partial}{\partial t}\mathcal{EE}\left(  r,t\right)  =\left[
\frac{p^{2}}{2m}+W(r)  \sin^{4}\theta+cK\cos^{2}\theta\right]
\mathcal{EE}\left(  r,t\right)   \label{A19}%
\end{equation}
with the initial condition
\begin{equation}
\mathcal{EE}(K,r,0)= \cos^{4}\theta f(K,r)  .
 \label{A20}%
\end{equation}

% ================================
\section{Bound states in three dimensions}\label{sec:app:bound_states_in_3D}
% ================================
In the main text we assume that the system is one-dimensional and the results are strictly valid only in this case. As in experimental setups only an approximate confinement to one dimension can be achieved we analyze the influence of higher dimensions. Therefore we consider the effective Hamiltonian corresponding to the three dimensional problem,
\begin{equation}\label{app:eq:H_eff_3D}
	H = -\frac{1}{2\abs{m}}\frac{\partial^2}{\partial r^2}-\frac{\vgroup}{k_\mathrm p}\left(\frac{\partial^2}{\partial x^2}+\frac{\partial^2}{\partial y^2}\right)+W(\abs{\vec r}),
\end{equation}
where $ \kp=\omegap/c $ denotes the carrier wave number of the probe field. In order to neglect the transversal kinetic energy terms
\begin{equation}
	\frac{\vgroup}{\kp}\left(\frac{\partial^2}{\partial x^2}+\frac{\partial^2}{\partial y^2}\right)
\end{equation}
compared to the longitudinal one, the following condition should be fulfilled,
\begin{equation}
	\frac{1}{2\abs{m}}\frac{1}{r_B^2}\gg 2\frac{\vgroup}{\kp}\frac{1}{w^2},
\end{equation}
where $ w $ denotes the probe beam waist. We can rewrite this as a condition for the  parameter $ \xi $, which yields
\begin{equation}\label{app:eq:dimensionalityCondition}
	\xi\gg \sqrt{\frac{\lambda_\mathrm{p}\Labs}{2\pi w^2}},
\end{equation}
i.e., imposing a lower bound on the interaction strength.

\end{document}